\documentclass[showpacs,prl,twocolumn]{revtex4}

\usepackage{latexsym}
\usepackage{amsfonts}
\usepackage{amssymb}
\usepackage{amsmath}
\usepackage{graphicx,color}

\begin{document}

\title{Electrical detection of spin pumping due to the precessing
magnetization of a single ferromagnet}

\author{M. V. Costache, M. Sladkov, S. M. Watts, C. H. van der Wal and B.
J. van Wees} \affiliation{Department of Applied Physics and
Materials Science Center, University of
Groningen,\\
Nijenborgh 4, 9747 AG Groningen, The Netherlands}

\date{\today}

\begin{abstract}
We report direct electrical detection of spin pumping, using a
lateral normal metal/ferromagnet/normal metal device, where a
single ferromagnet in ferromagnetic resonance pumps spin polarized
electrons into the normal metal, resulting in spin accumulation.
The resulting backflow of spin current into the ferromagnet
generates a d.c. voltage due to the spin dependent conductivities
of the ferromagnet. By comparing different contact materials (Al
and /or Pt), we find, in agreement with theory, that the spin
related properties of the normal metal dictate the magnitude of
the d.c. voltage.

\end{abstract}
\maketitle

Recent theoretical work in the field of spintronics
\cite{bauer02a,brataas,Xuhui} proposes to realize nanodevices in
which a so-called spin pumping mechanism is used for polarizing
electron spins in a normal metal (paramagnetic) region. Spin
pumping \cite{bauer02a} is a mechanism where a pure spin current
is emitted at the interface between a ferromagnet with a
precessing magnetization and a normal metal region. It is an
important new mechanism for controlling spins, since other
electronic methods based on driving an electrical current through
a ferromagnet (F)- normal metal (N) interface
\cite{jedema:Nat2001} are strongly limited by the so-called
conductance mismatch \cite{schmidt:2000}. Until now however, spin
pumping has only been demonstrated with thin multilayers, where it
appears as an enhanced damping of magnetization dynamics
\cite{Mizukami02,Urban01,Heinrich03,Lenz}.

\begin{figure}
\includegraphics[width=6cm]{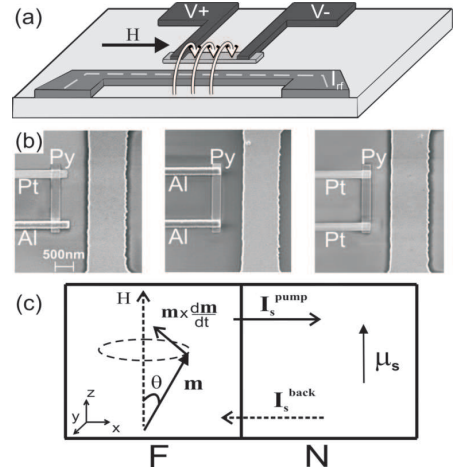}
\caption{(a) Schematic diagram of the device. On the lower side,
through the shorted-end of a coplanar strip a current $I_{rf}$
generates an rf magnetic field, denote by the arrows. The Py strip
in the center produces a d.c. voltage $\Delta V=V^{+}-V^{-}$. H
denotes the static magnetic field applied along the strip. (b)
Scanning electron microscope pictures of the central part of the
devices. (c) The F/N structure in which the resonant precession of
the magnetization direction $\mathbf{m}$ pumps a spin current
$\mathbf{I}^{pump}_{s}$ into N. The spin pumping builds up a spin
accumulation $\boldsymbol{\mu}_{S}$ in N that drives a spin
current $\mathbf{I}^{back}_{s}$ back into the F.} \label{str}
\end{figure}

In this Letter, we present spin pumping with a single nanomagnet
in an electronic device, in which it is directly detected as d.c.
electronic signal. The elementary mechanism is based on the
parametric spin pumping proposed by \cite{bauer02a}. As shown in
Fig. \ref{str}(c), a spin current
$\mathbf{I}_{s}^{pump}=(1/4\pi)\hbar g_{\uparrow\downarrow}
\mathbf{m}\times d\mathbf{m}/dt$ is pumped by (resonant)
precession of a ferromagnet magnetization into an adjacent normal
metal region. \textbf{m} is the magnetization direction and
$g_{\uparrow\downarrow}$ is mixing conductance \cite{bauer00}, a
material parameter which describes spin transport perpendicular to
\textbf{m} at the interface. Depending on the spin related
properties of the normal metal, two regimes exist. When the normal
metal is a good "spin sink" (in which spins relax fast), the
injected spin current is dissipated fast and this corresponds to a
loss of angular momentum and an increase in the effective Gilbert
damping of the magnetization precession
\cite{Mizukami02,Urban01,Heinrich03,Lenz}. However, in the limit
of the spin flip relaxation rate smaller than the spin injection
rate, a spin angular momentum builds up in the normal metal, i.e.
a spin accumulation $\boldsymbol{\mu}_{S}$ (difference between the
chemical potentials for spins up and down) exists in the normal
metal close to the interface \cite{brataas}. Due to electron
diffusion in the normal metal, the spin accumulation can diffuse
away from the interface and in principle, can be measured
electrically by using a second ferromagnet as a spin dependent
contact, placed at a distance shorter than the spin flip length
\cite{jedema:Nat2002,brataas}.

However, \citet{Xuhui} predicted a more direct way to detect it by
converting the spin accumulation into a voltage using the
precessing ferromagnet as its own detector, as illustrated in Fig.
\ref{str}(c). As a result of $\boldsymbol{\mu}_{S}$ a backflow
current goes back into F. The component parallel to \textbf{m} can
enter F and gives rise to a d.c. voltage due to the spin dependent
conductivities (bulk and interface) of the ferromagnet. Therefore,
in a device geometry where a ferromagnet is contacted with two
normal metal electrodes, any asymmetry between the two contacts
can result in a net d.c. voltage. The largest such asymmetry is
obtained when one of the metal electrodes is a "spin sink" (for
which we expect negligible d.c. voltage) such as Pt, while the
other has a small spin flip relaxation rate, such as Al. Here, we
describe precise, room-temperature measurements of the d.c.
voltage across a ferromagnetic strip contacted by Pt and Al
electrodes when the ferromagnet is in resonance. As control
devices, we also used contact configurations consisting of two Pt
electrodes and two Al electrodes. We found that the primary
contribution to the observed dc voltages comes from the Al
contact, thus allowing us to rule out spurious magnetoresistive
contributions.

Figure \ref{str}(a) shows a schematic illustration of the lateral
devices used in the present study. The central part of the device
is a ferromagnetic strip of permalloy (Py=$Ni_{80}Fe_{20}$)
connected at both ends to normal metals, Al and/or Pt ($V^{-}$ and
$V^{+}$ contacts). The devices are fabricated on a $Si/SiO_{2}$
substrate using e-beam lithography, material deposition and
lift-off. A 25 nm thick Py strip with 0.3x3 $\mu m^2$ lateral size
was e-beam deposited in a base pressure of 1x$10^{-7}$ mBar. Prior
to deposition of the 30 nm thick Al or/and Pt contact layers, the
Py surface was cleaned by Ar ion milling, using an acceleration
voltage of 500 V with a current of 10 mA for 30 sec, removing the
oxide and few nm of Py material to ensure transparent contacts. We
measured in total 17 devices (this includes 4 devices with a
modified contact geometry, described later). Different contact
material configurations are shown in Fig. \ref{str}(b).

We measured the d.c. voltage generated between the $V^{+}$,
$V^{-}$ electrodes as a function of a slowly sweeping magnetic
field ($H$) applied along the Py strip ($\textbf{z}$ axis), while
applying an rf magnetic field ($h_{rf}$) perpendicular to the
strip ($\textbf{y}$ axis). We have recently shown \cite{me} that a
submicron Py strip can be driven into the uniform precession
ferromagnetic resonance mode, using a small perpendicular rf
magnetic field created with an on-chip coplanar strip waveguide
\cite{Gupta} positioned close to Py strip (similar geometry as
shown in Fig. \ref{str}). For the used rf power level (9 dBm) an
rf current of $\approx12~mA$ \cite{cur} passes through the
shorted-end of the coplanar strip waveguide and creates an rf
magnetic field with an amplitude of $h_{rf}\approx1.6~mT$ at the
location of the Py strip. We could confirm that on resonance the
precession cone angle is $\approx5^{0}$ \cite{meAMR}.

To reduce the background d.c. offset and noise we adopted a
lock-in microwave frequency modulation technique. During a
measurement where static magnetic field is swept from $-400~mT$ to
$+400~mT$, the rf field is periodically switched between two
different frequencies and we measured the difference in d.c.
voltage between the two frequencies $\Delta
V=V(f_{high})-V(f_{low})$ using a lock-in amplifier. For all the
measurements the lock-in frequency is $17~Hz$ and the difference
between the two microwave frequencies is $5~GHz$.

\begin{figure}
\includegraphics[width=6.5cm]{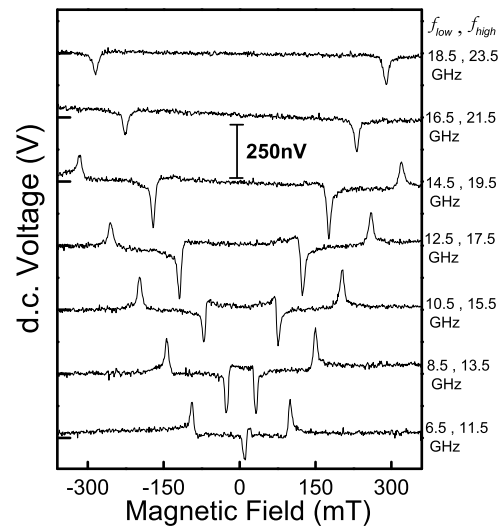}
\caption{The d.c. voltage $\Delta V$ generated by a Pt/Py/Al
device in response to the r.f. magnetic field plotted as a
function of the static magnetic field. The frequencies of the r.f.
field are as shown. The peaks (dips) correspond to resonance at
$f_{high} (f_{low})$. The data are offset vertically, for
clarity.} \label{fig2}
\end{figure}

\begin{figure}
\includegraphics[width=7cm]{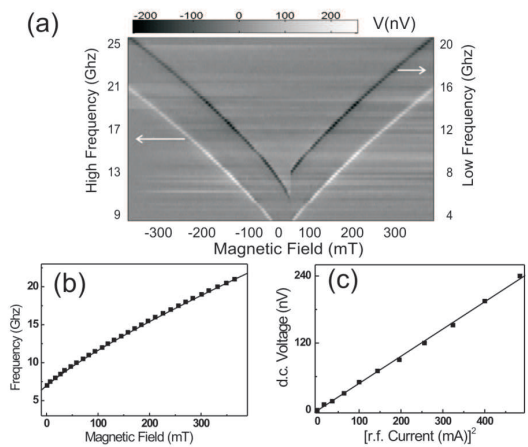}
\caption{(a) Gray scale plot of the d.c. voltage $\Delta V$,
measured function of static field for different high (low)
frequencies of the r.f. field from the Pt/Py/Al device
\cite{switch}. The dark (light) curves denote resonance at
$f_{low} (f_{high})$. (b) The static magnetic field dependence of
the resonance frequency of the Py strip (dots). The curve is a fit
to equation (1). (c) The amplitude of the d.c. voltage from a
Al/Py/Al device as a function of the square of the r.f. current,
at 13 Ghz and 139 mT (dots). The line shows a linear fit.}
\label{fig3}
\end{figure}

Figure \ref{fig2} shows the electric potential difference $\Delta
V$ from a Pt/Py/Al device. Sweeping the static magnetic field in a
range $-400~mT$ to $+400~mT$, a peak and a dip like signal are
observed at both positive and negative values of the static field.
Since we measured the difference between two frequencies, the peak
corresponds to the high resonant frequency ($f_{high}$) and the
dip to the low resonant frequency ($f_{low}$). For opposite sweep
direction the traces are nominally identical. We measured 8
devices with contact material Pt/Py/Al. The measured resonances
are all in the range $+100~nV$ to $+250~nV$. Notably, the d.c.
voltages are all of the same sign (always a peak for $f_{high}$),
meaning that for Pt/Py/Al devices, the Al contact at resonance is
always more negative than Pt contact.

First, we look at peak/dip position dependence of the rf
frequency. In Figure \ref{fig3}(a), the d.c. voltage in gray scale
is plotted versus static field for different high (low)
frequencies of the rf field. Figure \ref{fig3}(b) shows the
fitting of the peak/dip position dependence of the rf field
frequency (dotted curve) using the Kittel's equation for a small
angle precession of a thin-strip ferromagnet \cite{kittel}:
\begin{equation}
f=\frac{\gamma}{2\pi}\sqrt{(H+N_{\parallel}M_{S})(H+N_{\perp}M_{S})}
\label{Kittel}
\end{equation}
where $\gamma$ is the gyromagnetic ratio, $N_{\parallel}$,
$N_{\perp}$ are in plane (along the width of the strip) and out of
plane demagnetization factors and $M_{S}$ is the saturation
magnetization. The fit to this equation (see Fig. \ref{fig3}(b))
gives $\gamma=176~GHz/T$, and $N_{\parallel}\mu_{0}M_{S}=60~mT$,
$N_{\perp}\mu_{0}M_{S}=930~mT$, consistent with earlier reports
\cite{grundler_apl,me}. The fit confirms that the d.c. voltage
appears at the uniform ferromagnetic resonance mode of the Py
strip. Secondly, we measured the peak/dip amplitudes for different
values of the applied rf current \cite{cur} (Fig. \ref{fig3}(c)).
We observe here a linear dependence on the square of the rf
current. This is consistent with the prediction of the spin
pumping theory, as explained below.

\begin{figure}
\includegraphics[width=6.5cm]{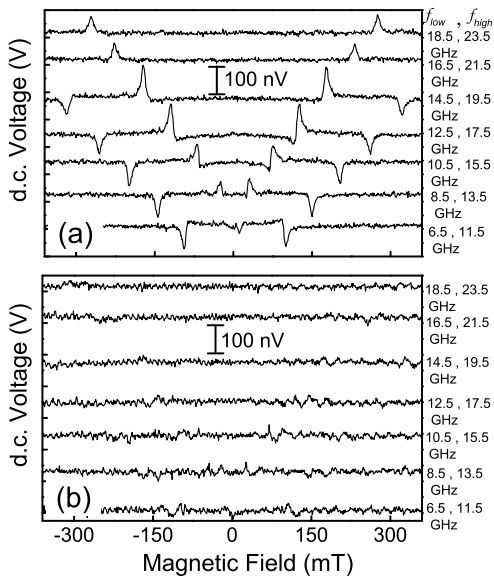}
\caption{The d.c. voltage $\Delta V$ generated across the Al/Py/Al
(a) and Pt/Py/Pt (b) devices as a function of the static magnetic
field. The frequencies of the r.f. field are as shown.}
\label{fig4}
\end{figure}

Further, we studied several control devices where the Py strip is
contacted at both ends by the same non-magnetic material Al (5
devices) or Pt (4 devices). Here we expected no signal because:
(i) Symmetry reasons, the voltages for identical interfaces are
the same (but opposite) and their contribution to $\Delta V$
cancels. (ii) Pt has a very short spin diffusion length, resulting
in a small spin accumulation, a small backflow and thus a lower
signal. The results from Al/Py/Al devices show smaller signals
than Pt/Py/Al devices, with a large scatter in amplitude and both
with positive and negative sign for the resonance at $f_{high}$.
Typical values for the 5 devices are $-100~nV$ (shown in Fig.
\ref{fig4}(a)), $+25~nV$, $+30~nV$, $+75~nV$ and $+110~nV$. In
contrast, all 4 Pt/Py/Pt devices exhibit only weak signals up to
$20~nV$ (with resonance signals barely visible, as in Fig.
\ref{fig4}(b)). We attribute the signals from Al/Py/Al devices to
the asymmetry of the two contacts (possible caused by small
variation in the contacts geometry and interface, see Fig.
\ref{str}(b)). Depending on the asymmetry, the signal therefore
have a scatter around zero. In the Pt/Py/Pt devices, independent
of possible asymmetry we expected and found very small signals. We
therefore conclude that the resonances measured with the Pt/Py/Al
devices arises mainly from the Al/Py interface (the Pt/Py/Al
devices have signals that are always positive, on average $\approx
+150~nV$, and with a scatter comparable in amplitude to that of
Al/Py/Al devices around zero).

In order to obtain the magnetization precession cone angle
$\theta$ \cite{meAMR} we performed anisotropic magneto resistance
(AMR) measurements. The measurements were carried out as before
(at $15.5~GHz$), but in addition a d.c. current (typically $50~\mu
A$) was sent trough the Py strip. Around the ferromagnetic
resonance an extra signal due to the AMR effect is measured (that
is linear with the applied d.c. current). For the AMR effect, the
change in the resistance (V/I) of the Py depends on the angle
between the d.c. current and the (time-averaged) direction of the
magnetization as $R(\theta)=R_{0}+\Delta R_{AMR}\cdot
cos^2(\theta)$, where $R_{0}$ is the resistance of the strip when
the magnetization is perpendicular on the direction of the current
and $\Delta R_{AMR}$ is the change in the resistance between
parallel and perpendicular directions of the magnetization. From
the amplitude of the d.c.-current-induced contribution to the
resonance, and using a measured AMR value of $2\%$ for our Py
strip, we obtained $\theta$ to be $\approx 5^{0}$. This value is
consistent with the value $\theta\approx h_{rf}/(\alpha
M_{S})=6^{0}$ calculated using the solution of the linearized LLG
equations \cite{LLG}. Here a damping parameter of $\alpha = 0.012$
results from fitting the AMR voltage shape at resonance with the
frequency dependent value of $\theta ^2$.

We also checked that the observed resonances do not arise from
rectification effects that can result from time dependent AMR. We
analyzed \cite{meAMR} that due to the capacitive and inductive
coupling between the CSW and the Py strip an rf current can be
induced in the detection circuit. In combination with a time
dependent AMR (which can contain a component with rf frequency
$\omega$) this can lead to a rectification effect where a d.c.
voltage is created. To rule out a possible contribution to the
measured resonance signals, we have studied 4 devices similar to
Fig. \ref{str}(b), but now with contacts at the ends of the Py
strip (extending along the $\textbf{z}$ axis) \cite{new_gr}. We
found no significant difference in the observed d.c. voltages
between this geometry and that shown in Fig. \ref{str}(b).
Further, we misaligned the direction of the static field with
respect to the Py strip long axis, and only found significant
contributions from rectification effects at offset angles higher
than 5 degrees. This rules out that small offset angles caused
significant effects in the present study.

We now analyze the results. As can been seen in Fig. \ref{str}(c),
due to a time dependent magnetization direction \textbf{m} the
pumped spin current has a constant component in the \textbf{z}
direction and oscillating components in \textbf{x} and \textbf{y}
directions. Here, we are interested in the description for small
$\theta$. We do not take into account spin relaxation in N because
for small $\theta$ the spin relaxation in F will be the dominant
mechanism for controlling the magnitude of the d.c. voltage. The
dynamics of \textbf{x} and \textbf{y} components of spin are
controlled by the length scale $l_{\omega}=\sqrt{D_{N}/\omega}$
($D_{N}$ is the diffusion coefficient in the N and $\omega$ is the
precessional frequency) which describes the length scale over
which the averaging of \textbf{x} and \textbf{y} components
occurs. In our experiment $l_{\omega}$ of Al is order of 200-300
nm. This means that in principle we have to fully model spin
dynamics in this region. However, we follow \citet{Xuhui} to get a
qualitative estimate of the effect. It is assumed that \textbf{x},
\textbf{y} components are fully averaged and therefore zero and
the remaining \textbf{z} component is constant and along the
static magnetic field direction \cite{brataas}. Also, for small
$\theta$ the component of spin accumulation $\boldsymbol{\mu}_{S}$
parallel to \textbf{m} is approximatively equal to $\mu_{S}$. This
component can diffuse back into F and give rise to a d.c. voltage
due to spin dependent conductivities. Thus a voltage of
$p_{\omega}\mu_{S}$ will be generated across the interface. For
small angle precession this results in \cite{Xuhui}

\begin{equation}
V_{dc} =\frac{p_{\omega}
g_{\omega}^{\uparrow\downarrow}}{2e(1-p_{\omega}^{2})
g_{\omega}}\theta^{2} \hbar \omega
\end{equation}

where $p_{\omega}=(g_{\omega}^\uparrow-g_{\omega}^\downarrow)/
(g_{\omega}^\uparrow+g_{\omega}^\downarrow)$ and
$g_{\omega}=g_{\omega}^\uparrow+g_{\omega}^\downarrow$ with
$g_{\omega}^\uparrow(g_{\omega}^\downarrow)$, the spin up (down)
effective conductances of the Py/Al interface \cite{cond}. The
quadratic dependence of $\theta$ of eq. (2) is in agreement with
the experimental data (see Fig. \ref{fig3}(c)). Having determined
a precession cone angle $\theta\approx5^{0}$, using
$p_{\omega}\approx 0.2$, an FMR frequency $\omega=10^{11}~s^{-1}$
and $g_{\omega}^{\uparrow\downarrow}/g_{\omega}\approx 1$, we find
$V_{dc}\approx 100~nV$, in reasonable agreement with the
experimental results.

In summary, we have measured a d.c. voltage due to the spin
pumping effect, across the interface between Al and Py at
ferromagnetic resonance. We find that the devices where the Al
contact has been replaced by Pt show a voltage close to zero, in
good agreement with theory. Although the prediction of the spin
pumping model fairly well agrees with the observed d.c. voltage
values, a more detailed description is required that would include
the elliptical precession motion of the ferromagnet's
magnetization as well as the spin dynamics in N \cite{Xuhui02}.

This research was supported by the 'Stichting voor Fundamenteel
Onderzoek der Materie (FOM)'.

\bibliographystyle{apsrev}
\bibliography{ref}

\begin{thebibliography}{22}
\bibitem[{cur()}]{cur}
\bibinfo{note}{The 9 dBm applied rf power was converted into a rf current
  assuming $50\Omega$ load impedance.}
\bibitem[{swi()}]{switch}
\bibinfo{note}{The irregular feature, only occurring near $H=0$ mT is due to
  magnetization switching with hysteresis.}
\bibitem[{con()}]{cond}
\bibinfo{note}{$g_{\omega}^\uparrow$ ($g_{\omega}^\downarrow$) are the spin-up
  (down) interface conductances is series with a conductance of the bulk nomal
  metal over a length scale of $l_{\omega}$ and
  $g_{\omega}^{\uparrow\downarrow}$ is the mixing conductance evaluated over a
  length scale of $l_{\omega}$}.
\end{thebibliography}

\end{document}